\documentclass[prl,amsmath,amssymb]{revtex4}
\usepackage{graphicx}

\newcommand{\unit}[1]{\ensuremath{\,\mathrm{#1}}}
\newcommand{\um}{\ensuremath{\unit{\mu m}}}

\newcommand{\expect}[1]{\ensuremath{\left\langle #1 \right\rangle}}
\newcommand{\fig}[1]{figure~\ref{#1}}
\newcommand{\Fig}[1]{Figure~\ref{#1}}

\newcommand{\ddt}{\ensuremath{\frac{\mathrm{d}}{\mathrm{d}t}}}
\newcommand{\DD}{\ensuremath D}
\newcommand{\Dsp}{D_{\rm sp}}
\newcommand{\Ddp}{D_{\rm dp}}

\newcommand{\csig}{\sigma^{\dag}}

\newcommand{\sig}{\sigma}
\newcommand{\msig}{\left<\sig\right>}

\newcommand{\aop}{a}
\newcommand{\maop}{\left<\aop\right>}

\newcommand{\caop}{a^{\dag}}

\begin{document}

\title{Light force fluctuations in a strongly coupled atom-cavity
system}

\author{T. Puppe}
\author{I. Schuster} \affiliation{Max-Planck-Institut f\"ur Quantenoptik,
Hans-Kopfermann-Str. 1, 85748 Garching, Germany}
\author{P. Maunz} \altaffiliation{present address: FOCUS Center and
Department of Physics, University of Michigan, Ann Arbor, Michigan
48109, USA}
\author{K. Murr}
\author{P.W.H. Pinkse}
\author{G. Rempe}
\affiliation{Max-Planck-Institut f\"ur Quantenoptik,
Hans-Kopfermann-Str. 1, 85748 Garching, Germany}

\date{\today, PREPRINT}

\begin{abstract}
Between mirrors, the density of electromagnetic modes differs from
the one in free space. This changes the radiation properties of an
atom as well as the light forces acting on an atom. It has profound
consequences in the strong-coupling regime of cavity quantum
electrodynamics. For a single atom trapped inside the cavity, we
investigate the atom-cavity system by scanning the frequency of a
probe laser for various atom-cavity detunings. The avoided crossing
between atom and cavity resonance is visible in the transmission of
the cavity. It is also visible in the loss rate of the atom from the
intracavity dipole trap. On the normal-mode resonances, the dominant
contribution to the loss rate originates from dipole-force
fluctuations which are dramatically enhanced in the cavity. This
conclusion is supported by Monte-Carlo simulations.
\end{abstract}

\maketitle

\section{Introduction}

``As a subject of continuing interest light stands the test of time
very well'', Knight and Allen wrote in 1983~\cite{KnightAllen}.
Today, this is as true a statement as it was 24 years ago. There is
hardly a laboratory without a laser and the quantum properties of
light continue to rouse researchers' interest. The ideal environment
to study the interaction between light and matter at the level of
single quanta is an optical cavity. In the strong-coupling regime,
the periodic exchange of the excitation between the atom and the
cavity is faster than all loss rates in the system, and the atom and
cavity resonances exhibit an avoided crossing, as predicted by the
Jaynes-Cummings model (JCM) \cite{Jaynes63,Shore93}. The resulting
splitting, which exists in the limit of weak driving, is referred to
as the ``vacuum-Rabi'' splitting. Indeed, the vacuum-Rabi splitting
has become a benchmark measurement for strong
coupling\cite{Maunz05,Boca04}.

Because of the coupling, the bare states of the atom and the cavity
mode mix to form new  energy eigenstates, the so-called ``dressed
states''. In a typical experiment the normal-mode splitting is
measured in the excitation spectrum of the cavity. Since the
transmission of the cavity is directly proportional to the cavity
excitation, such a measurement is straightforward. From a
theoretical point of view, measuring the atomic excitation via
spontaneous emission is equivalent. The combination of these two
measurements would complete the information on the excitation of the
dressed-state constituents. Experimentally, however, the
spontaneously emitted photons are not easily accessible. Recognizing
that the recoils of spontaneous emission lead to heating of the
atom, a naive picture predicts that the atomic excitation can be
observed by monitoring the heating of an atom as a function of the
laser frequency.

The normal-mode spectrum has been measured in both the transmission
of the cavity and the loss rate of the atom from an intracavity
trap~\cite{Maunz05}. It turns out that the loss spectrum cannot be
explained by the spontaneous emission alone. This shows that the
naive picture is incomplete because there are more light forces
acting on an atom in a cavity than the one due to spontaneous
emission. For instance, in addition to the (conservative) dipole
force, theory predicts velocity-dependent light forces in a cavity,
which are exploited in cavity cooling
~\cite{Mossberg91,Horak97,Vuletic00,Maunz04}. The cavity is
indispensable in this scheme, as it provides a necessary delay in
the response of the field to the presence of the atom. The cavity
also modifies the {\em fluctuations} of the light force, which
increase the kinetic energy of an atom and thus open up an
additional escape mechanism for an atom trapped inside the cavity.
The rate at which the kinetic energy is increased is quantified by
the momentum diffusion coefficient. Previous calculations
~\cite{Horak97,Hechenblaikner98,Domokos00,Fischer01,Murr03,Domokos03}
have shown that dipole-force fluctuations can dramatically enhance
momentum diffusion inside a cavity. A recent general derivation
identifies a term in the diffusion coefficient which originates from
the cavity dissipation\cite{MurrPRA06}. Compared to the
corresponding free-space situation, this leads to a large increase
of the momentum diffusion.

The present paper gives a detailed account of the role of
spontaneous emission and of the dipole-force fluctuations in a
cavity QED system with single trapped atoms. Experimental
data~\cite{Maunz05} are compared with Monte-Carlo simulations in
which the different contributions to the light force can be
evaluated individually. After a short introduction of relevant
theoretical aspects of the atom-cavity system we show the
experimental results and compare the data with detailed numerical
simulations.

\section{Vacuum-Rabi splitting}
\label{normalmode_AvoidedCrossing}

The closed quantum system composed of a single atom and a single
mode of the light field was described by Jaynes and Cummings
\cite{Jaynes63} and can be solved analytically. The Hamiltonian of a
single mode of frequency $\omega_{cav}$ with creation operator
$\caop$ in interaction with a two-state atom of frequency
$\omega_{at}$ with lowering operator $\sig=|g\rangle\langle e|$ in
the dipole and rotating-wave approximation reads:
\begin{eqnarray}
H_{JC}/\hslash &=& \omega_{cav}\caop\aop+\omega_{at}\csig\sig + g(\aop\csig+\caop\sig) \ ,
\end{eqnarray}
where $g$ is the atom-cavity coupling constant. The two eigenstates
$|+\rangle$ and $|-\rangle$ of the first excited doublet of the
combined atom-cavity system are superpositions of the atomic ground
state with one intracavity photon $|g,1\rangle$ and the atomic
excited state with zero photons $| e,0 \rangle$:
\begin{eqnarray}
\label{DressedStatesPos}%
|+\rangle &= \sin \theta | g,1 \rangle + \cos \theta | e,0 \rangle \\
\label{DressedStatesNeg}%
|-\rangle &= \cos \theta | g,1 \rangle -
\sin \theta | e,0 \rangle\ .
\end{eqnarray}
These ``dressed states'' are obtained by a rotation in Hilbert space
from the bare state basis $\{\, | g \rangle,| e \rangle\}$ to the
dressed state basis $\{\, | + \rangle,| - \rangle\}$ by the mixing
angle $\theta$. The mixing angle ($0\le\theta<\pi/2$) is defined by
$\tan 2 \theta=-2g/\Delta$. It depends on the detuning between the
cavity and the atomic resonance, $\Delta=\omega_{cav}-\omega_{at}$.
For the detunings $\Delta=(-\infty,0,\infty)$ the mixing angle is
$(0,\pi/4,\pi/2)$. By convention~\cite{CCTAtomPhoton}, the energy of
the state $|+\rangle$ is larger than that of the state $|-\rangle$.
In the closed system, the dressed states have well-defined
eigenenergies.

The open quantum system is subject to loss of photons from the
cavity and spontaneously emitted photons from the atom and can be
solved analytically in the limit of weak atomic
excitation~\cite{Carmichael91,Hechenblaikner98}. To replenish lost
photons, the system must be pumped by an incident light field,
called probe laser. Since in our case only the cavity mode is
excited by the probe laser, the excitation of a dressed state is
proportional to the contribution of the cavity state to the dressed
state and to the probe laser power. In the open quantum system the
dressed states have finite energy uncertainties and, hence, finite
linewidths. The excited bare cavity state and the excited atomic
state have linewidths of $2\kappa$ and $2\gamma$, respectively. In
general, the linewidth of the dressed states will be given by a
weighted average of $2\kappa$ and $2\gamma$ determined by the mixing
angle. For large detunings the mixing angle is small such that the
dressed states converge to the bare states with the corresponding
linewidth.

\begin{figure}
\includegraphics[width=0.75\textwidth]{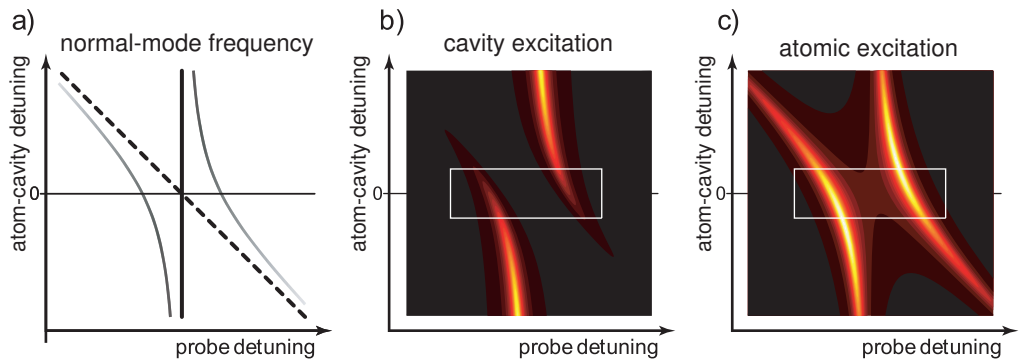}
\caption{Normal modes as a function of atom-cavity and (probe)
laser-cavity detunings. a) The coupling of the atom (frequency
determined by the straight dashed line) and the cavity (vertical
solid line) generates two new eigenstates which are superpositions
of atomic and cavity states (curved lines). For vanishing detuning
between atom and cavity the dressed states reach the minimum
separation of $2g$, the vacuum-Rabi splitting. b) Cavity excitation
as a function of atom-cavity and probe detunings. c) Atomic
excitation as a function of atom-cavity and probe detunings. The
cavity excitation on the dressed-state resonances depends more
strongly on the mixing angle than the atomic excitation, as only the
cavity part of the dressed state is excited by the probe laser. In
b) and c) the frequency range covered by the experiment (see
Fig.~\ref{fig_avoidedcrossingdata}) is indicated by the white
rectangle.}%
\label{fig_avoidedcrossingtheory}
\end{figure}

\Fig{fig_avoidedcrossingtheory} b) shows the calculated cavity
transmission as a function of the probe detuning and of the
atom-cavity detuning. The probe detuning is indicated with respect
to the cavity resonance. The transmission of the cavity is
proportional to the number of photons in the cavity. Therefore the
height of the two normal modes in the transmission spectrum is
proportional to $\sin^2\theta$ and $\cos^2\theta$, respectively. For
zero detuning between atom and cavity, the mixing angle is $\pi/4$
and the transmission at both normal-mode frequencies is equal. In
this case, the distance between the two normal-mode frequencies
reaches its minimum value of $2g$, the vacuum-Rabi splitting.

In contrast to the transmission, the calculated atomic excitation
spectrum (\fig{fig_avoidedcrossingtheory} c) shows a different
dependency of the peak height on the mixing angle. As here again the
cavity is excited whereas now the atomic excitation is plotted, the
height of both normal modes is proportional to $\sin\theta\,
\cos\theta$ and hence only exhibits a weak dependency on the
atom-cavity detuning.

\section{The experiment}
\label{experiment}

Our experimental setup and the measurement procedure used to observe
the normal modes of a single atom strongly coupled to a cavity were
described in detail elsewhere \cite{Maunz05}. In short, laser-cooled
$^{85}$Rb atoms are injected into a high-finesse cavity with
velocities below 10\,cm/s. The cavity has a finesse of
$4.4\times10^5$, a mode waist of 29\,\um\ and a length of 122\,\um.
It is near resonant with the $5^2S_{1/2}F=3,m_F=3 \leftrightarrow
5^2P_{3/2}F=4,m_F=4$ transition of $^{85}$Rb at 780.2\,nm. The
maximum coupling constant for this transition is
$g=2\pi\times16\,$MHz. A weak red-detuned (785\,nm) intracavity
dipole field guides the atom into its antinodes. The dipole mode has
two nodes fewer than the cavity-QED probe mode, so that the
antinodes of the two modes coincide midway between the cavity
mirrors. For the probe laser on resonance with the cavity, the
presence of an atom leads to a distinct drop of the cavity
transmission. When the transmission drops below a preset threshold,
an atom must have entered a region where the antinodes of the
trapping and probe fields overlap. The power of the dipole field is
suddenly increased, trapping the strongly coupled atom with close to
100\% efficiency. The dipole trap induces a position-dependent
differential Stark shift between the ground and excited states of
the atom, so that the detuning of the probe laser with respect to
the atom becomes a function of the dipole laser power and of the
position. After trapping the atom, the probe frequency and power is
adjusted to perform the desired spectroscopy. For probe frequencies
close to the normal modes, the dissipative light forces are strongly
heating. These frequencies cannot be avoided when scanning across
the spectrum. Therefore, measurement windows of 0.1\,ms duration
were interwoven with 0.5\,ms long time windows in which the
detunings are set for optimal cavity cooling~\cite{Maunz04},
relocalizing the atom in the axial direction and obtaining
information on the atom-cavity coupling for further data processing.

\begin{figure}%
\includegraphics[width=0.75\textwidth]{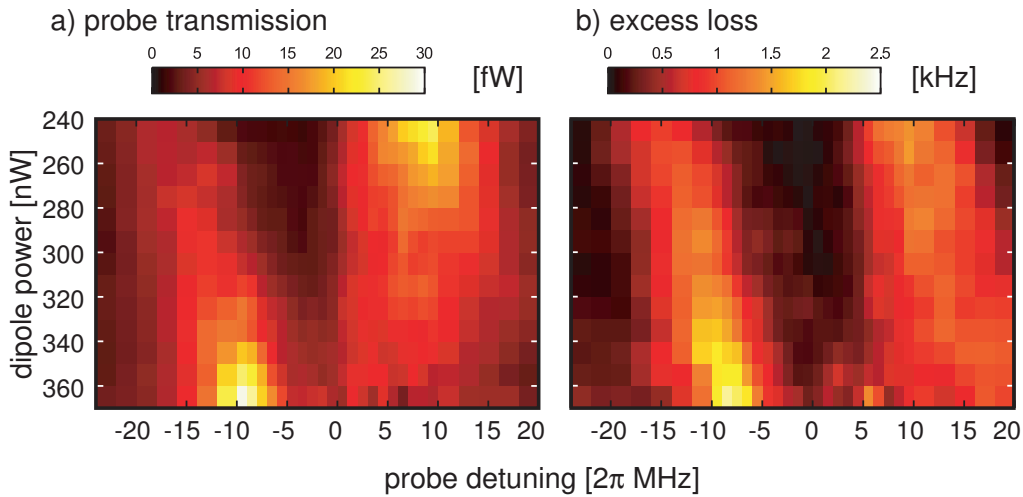}
\caption{Experimentally measured cavity transmission a) and excess
loss rate b) as a function of dipole power and probe detuning.
Tuning the dipole power changes the Stark shift induced by the
trapping field and results in a change of the effective atom-cavity
detuning. The two normal modes are clearly visible in the
transmission. They show the expected dependence on the detunings.
Compared to the transmission measurement, the observed excess loss
rate has a weaker dependence on the atom-cavity detuning. In both a)
and b) the broadening of the right normal mode with respect to the
left one is a consequence of the atomic motion, as explained in the
text. The data shown in figure \ref{fig_lossspectra} are also
included in b). } \label{fig_avoidedcrossingdata}
\end{figure}

In \fig{fig_avoidedcrossingdata} a), the cavity transmission is
plotted as a function of the probe detuning for different trap
depths given by the dipole laser power as measured in transmission.
Only observation windows were used where the neighbouring cooling
intervals indicated that the atom was well localized, ``qualifying''
those time windows where the atom was strongly coupled. For each
setting of the dipole power, i.e., a horizontal line in the figure,
two distinct resonances are observed when the probe detuning is
scanned. As can be seen, the positions of the peaks vary with the
dipole power. For small dipole power, the bare atomic resonance has
a lower frequency than that of the cavity, and the asymmetric
normal-mode spectrum shows the weak atom-like peak for negative
probe detunings and a stronger cavity-like peak for small positive
detuning of the probe light. For high dipole power, the effective
atomic resonance is found at frequencies higher than the bare cavity
resonance. The cavity-like peak, which is now the one at lower probe
detuning, is again stronger. Hence, the data show the avoided
crossing of the dressed states and the strong dependence of the
transmission on the atom-cavity detuning is clearly visible.

In contrast to the cavity transmission, the atomic excitation cannot
be measured directly in our experiment. We can, however,
experimentally determine the loss rate from the trap. From the
observed mean storage time the {\em excess} loss rate is derived,
defined as the surplus loss rate induced by probing the system. It
is shown in \fig{fig_avoidedcrossingdata} b). The simulations in the
next paragraph show that the excess loss rate on the normal-mode
resonances is dominated by heating generated by fluctuations of the
dipole force. As will be discussed, it is actually because of the
large enhancement of the dipole fluctuations in a cavity QED
environment that the normal modes are clearly visible in the excess
loss rate.

\section{Simulations}
\label{simulations}%

To understand the details of the observed transmission spectra, in
principle only the distribution function of the atomic position is
required to determine the measured line shapes. The spatial
distribution is governed by light forces, which can be calculated:
our experiment operates at low saturation, where the atomic
excitation is small ($P_e\!\!\approx\!\!|\msig|^2\ll 1$). Here,
analytical expressions for the expectation values of the dipole
force, the velocity-dependent force and the momentum diffusion
coefficient are available, even in the presence of the dipole trap,
which gives rise to nontrivial changes of the light forces. One
might therefore expect that the position distribution function can
be predicted in a straightforward manner. This is not the case. The
expressions for the forces depend strongly on the position of the
atom and the light forces therefore determine the atomic motion in a
very complex way. Moreover, the loss spectra depend even stronger on
the details of the motion, and a position distribution function will
not be sufficient to accurately predict the escape probability of an
atom from the trap. This is the reason why we performed a Monte
Carlo simulation and calculated many trajectories of single
(point-like) atoms in three dimensions by integrating the Langevin
equations of motion. The trajectory is determined by the frequencies
of the atom, cavity and laser and the intensities of the
standing-wave trapping and probe fields and their geometry. In order
to model the experiment in detail, a single atom is injected
transversally into the cavity from below. In the centre of the
cavity the antinodes of the guiding and probe fields coincide. As in
the experiment, the atom is thus guided into a region of strong
coupling. For the parameters of the experiment, the presence of an
atom leads to a distinct drop of the cavity transmission. When the
simulated transmission drops below a threshold, the power of the
intracavity dipole laser is increased to capture the atom. After
trapping the atom, the sequential probing and cooling intervals are
precisely simulated according to the experimental procedure.

\label{numerics_Forces}

Fluctuations of the dipole force increase the width of the momentum
distribution. This is quantified by the momentum diffusion
coefficient $2D = \ddt \left( \expect{\mathbf{p}^2} -
\expect{\mathbf{p}}^2 \right)$. The expectation value of this
diffusion coefficient can be calculated analytically in the regime
of low saturation and in lowest order of the atomic velocity. The
result can be expressed as the sum of a spontaneous emission term
and a dipole force fluctuation term:
\begin{equation}
\label{DatTot} \DD=\Dsp+\Ddp.
\end{equation}
The spontaneous emission term is directly proportional to the
excitation of the atom, $P_e$:
\begin{equation}
\label{Dsp} \Dsp=(\hslash k)^2\gamma P_e.
\end{equation}
As shown in Ref.~\cite{MurrPRA06}, the dipole fluctuation term,
$\Ddp$, can be separated into two distinct contributions:
\begin{equation}
\label{Ddp} \Ddp=|\hslash\nabla\msig|^2\gamma +
|\hslash\nabla\maop|^2\kappa,
\end{equation}
where the mean coherence $\msig$ is to be evaluated within the
cavity setting and $\maop$ is the mean amplitude of the cavity
field. The first term in $\Ddp$ can be interpreted as caused by a
fluctuating atomic dipole coupled to a non-fluctuating light field.
The second term in $\Ddp$ can be interpreted as caused by a
fluctuating cavity field coupled to a non-fluctuating atomic dipole.
The expression for $D$, Eqs.~(\ref{DatTot},\ref{Dsp},\ref{Ddp}), is
valid regardless of the value of the coupling $g$, the only
condition is low saturation such that the internal state of the atom
can be treated as a harmonic oscillator.

The two contributions to $\DD$ differ in their spatial properties,
which requires some care in the implementation into the simulation.
The momentum diffusion due to the fluctuation of the dipole force is
proportional to the square of the gradient of the atom-cavity
coupling strength, which is about $50$ times larger in axial
direction than in the radial direction. Thus, the resulting heating
is mainly directed along the cavity axis. In the simulation, the
axial momentum of the atom is changed in each time step, the sign of
the change chosen randomly with equal probability. The emission of a
photon from the atom imprints the photon recoil of $\delta p=\hbar
k$ on the motion of the atom. The spatial distribution of scattered
photons and thus also of photon recoils is given by the emission
characteristics of the polarized excited state of the atom. For
$^{85}$Rb atoms in the $F=3, m_F=3$ electronic ground state, which
are excited by circularly polarized light, $2/5$ of the momentum
diffusion acts in direction of the cavity axis, which coincides with
the quantization axis, and $3/10$ points along each of the two
orthogonal radial directions. This diffusion mechanism is included
in the simulation by applying random momentum kicks. The momentum
kicks which are used to generate the stochastic force reproduce the
first moment of the spreading of the momentum distribution, the
diffusion coefficient. For the numerical integration of the
equations of motion a Runge-Kutta algorithm with adjustable step
size is used. For stochastic processes, a highly uniform
pseudorandom number generator is used, the ``Mersenne
Twister''\cite{Matsumoto98}.

In the experiment, the storage time in the trap without probe light
is found to be limited by parametric heating. This heating is caused
by technical fluctuations of the frequency difference between the
trapping laser and the cavity resonance, resulting in intensity
fluctuations of the intracavity field. A theoretical investigation
\cite{Savard97} shows that the heating rate is proportional to the
spectral noise power density of the intracavity intensity at twice
the trap frequency. In the simulation, this heating mechanism is
modelled by randomly changing the intensity of the trapping field in
each time step $\delta t$. This leads to a white spectrum of the
noise power density for frequencies up to $1/2\delta t$. The
intensity in each time step is chosen according to a Gaussian
distribution with mean value equal to the average intensity. The
width of the distribution is adjusted to reproduce the
experimentally observed storage time of an atom in the dark trap,
without probe light. It is verified that for these fixed parameters,
the model reproduces the observed storage times for different probe
powers~\cite{Maunz04}.

\section{Light-force fluctuations in simulation and experiment}

Results of the simulation for the transmission of the cavity were
shown in Ref.~\cite{Maunz05}. They agreed well with the experimental
data if the power of the dipole laser was reduced by about $30\%$
with respect to the intracavity power estimated from the measured
cavity transmission. This could be explained by different
transmission coefficients of the two cavity mirrors, which would
lead to an error in the calibration of the intracavity light
intensity. For consistency, the power of the probe light in the
simulation is reduced by the same amount.

\begin{figure}
\includegraphics[width=0.75\textwidth]{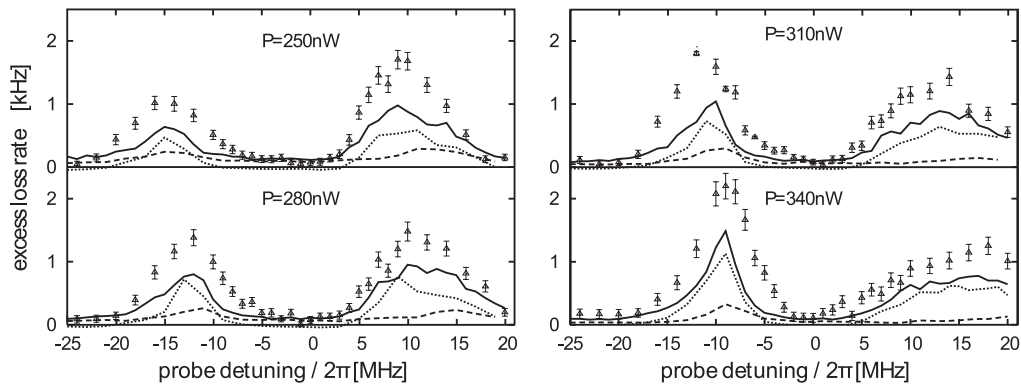}
\caption{Observed excess-loss rate of trapped atoms as a function of
the probe detuning (triangles). On average each point includes the
data from between $35$ and $1000$ atoms. The lines show results of
the Monte Carlo simulation including spontaneous emission only
(dashed lines), dipole fluctuations only (dotted lines), and the
contribution of both heating mechanisms (solid lines). }
\label{fig_lossspectra}
\end{figure}

Results of the simulation for the excess loss are shown in
\fig{fig_lossspectra}, together with the data \cite{Maunz05}. The
loss spectra also show two well-resolved peaks at the normal-mode
resonances. The left peak is observed to be narrower than the right
peak. This is explained by the position-dependent Stark shift: if an
atoms oscillates in its dipole trap, it will experience different
coupling strengths and different Stark shifts. Our trapping method
ensures that atoms are stored in an antinode of the dipole trap
overlapping with the antinodes of the cavity-QED field. Therefore,
an atom closer to the centre of its local trap will experience a
larger Stark shift and will simultaneously be stronger coupled. A
stronger coupling will make the splitting between the normal modes
larger; a stronger Stark shift will shift the entire spectrum to the
right (towards larger probe detunings). Hence, for the left peak
these two effects are opposed, while for the right peak they add up,
which explains the fact that the left one is narrower than the right
one. The results of the simulation reproduce this effect and predict
the line shapes in the loss spectrum well. The residual differences
in amplitude are considered small in view of the fact that absolute
loss rates are calculated without free parameters.

We now analyse the role of the light forces and their contribution
to the features in the spectrum. In this respect it is important to
realize that cavity cooling (either during probing or during the
cooling intervals) only cools the axial motion in our setup. Hence,
an atom is most likely to escape in radial direction after it is
heated up by spontaneous emission recoils. The heating due to
fluctuations in the amplitude of the dipole trapping field is
strongest in the axial direction, where the gradients of the fields
are largest. The heating in this direction is compensated by cavity
cooling, also working in the axial direction. This leads to the
hypothesis that the normal modes show up in the loss spectra because
on the normal-mode resonances the excitation is large and
spontaneous emission will heat up the atom in the transverse
direction, where it will finally escape the trap. In the following
we will see that this hypothesis is wrong.

In order to test our hypothesis, we exploit the possibilities of the
simulation. In the simulation one has complete information about the
atomic trajectory. This allows for example to test the role of the
transverse heating due to $\Dsp$ by identifying the direction in
which an atom escaped. Another advantage of the simulation is that
it is possible to enable or disable individual light forces in a
particular simulation run, and in this way determine the
contributions of the individual forces. In \fig{fig_lossspectra}
simulated loss spectra considering only heating from spontaneous
emission ($\Dsp>0, \Ddp=0$) are compared with simulations taking
into account fluctuations of the dipole force only ($\Dsp=0,
\Ddp>0$) and a simulation including both diffusion mechanisms
($\Dsp>0, \Ddp>0$). As expected, the results with only the
spontaneous emission heating (dashed line) show a normal-mode
spectrum. However, in addition to the difference in overall
amplitude, also the shape of the simulated spectra differs from the
measured loss spectrum. This falsifies the above hypothesis and one
is forced to conclude that other heating mechanisms must play an
important role.

The results of the simulation with only the dipole fluctuations
enabled (dotted line) show a normal-mode spectrum, which fits the
data better than the simulation with only the spontaneous emission
enabled. This result hints that on the normal-mode peaks the axial
heating is compensated insufficiently. The particle can then escape
in axial direction. This can be investigated in the simulation with
all forces (solid line), which describes the data best. The results
show that for zero and large probe detunings, spontaneous emission
dominates the losses. Here an atom leaves the cavity with high
probability in the transverse direction, the direction in which
heating is not compensated by cooling. If the probe light is
resonant with one of the normal modes, momentum diffusion induced by
dipole fluctuations dominates and causes more than two thirds of the
loss rate. In the simulation it is indeed observed that on the
normal-mode peaks, a significant amount of the atoms escape in the
axial direction. The conclusion is that on the normal modes, the
larger momentum diffusion is only partly compensated by cavity
cooling in the neighbouring cooling intervals.

The data show that the cavity-enhanced momentum diffusion is a real
and important process that cannot be neglected in experiments with
high-finesse cavities. It should be emphasized that this finding is
in accordance with theory: momentum diffusion on the symmetric
normal-mode resonances for an atom in our cavity is about 50 times
larger than in free space for the same intensity of a standing wave
probe field~\cite{enhancementfactor}. Between the normal-mode
resonances, the contribution of this term is close to zero; for
probe frequencies far detuned from the normal modes, the term falls
off quadratically. It is this remarkably large enhancement of $\Ddp$
at the normal modes which makes the normal modes clearly visible in
the excess loss rate.

\section{Conclusion}
\label{conclusion} In conclusion, we have observed the avoided
crossing between the dressed states of a strongly coupled
atom-cavity system. The crossing is clearly visible in both the
transmission and in the loss rate. Comparison of the latter with
simulations allows to deduce the relative importance of different
heating mechanisms and gives evidence for strong cavity-enhanced
heating on the normal-mode resonances, an effect which permits the
clear observation of the normal modes in the loss rate in the first
place. It should be emphasized that on the normal-mode resonances
the dominating heating originates not only from the enhanced atomic
dipole fluctuations, but also from enhanced fluctuations of the
cavity field, which significantly exceed the fluctuations in
free-space fields for the same intensity.


\begin{thebibliography}{12}
\bibitem{enhancementfactor}
The cavity enhancement of the momentum
diffusion~\cite{Horak97,Hechenblaikner98,Fischer01,Murr03} on the
symmetric normal-mode resonances reduces to $8C/(1+\gamma/\kappa)$.
With the cooperativity $C=g_{\rm eff}^2/(2\gamma\kappa)$ and an
effective coupling strength of $g_{\rm eff}=13\times 2\pi\,$MHz
obtained from the simulations, we find an enhancement of 51.
\end{thebibliography}
\end{document}